# Nature of the Monoclinic to Cubic Phase Transition in the Fast Oxygen Ion Conductor $La_2Mo_2O_9$ (LAMOX)

Lorenzo Malavasi[a], HyunJeong Kim[b], Simon J. L. Billinge[b], Thomas Proffen[c], Cristina Tealdi[a], Giorgio Flor[a]

[a] *Department of Physical Chemistry and Department of Pavia of IENI/CNR, University of Pavia, Pavia, 27100-Italy.*

[b] *Department of Physics and Astronomy, Michigan State University, East Lansing, Michigan 48824-1116, USA.*

[c] *Los Alamos National Laboratory, LANSCE-12, MS H805, Los Alamos, NM, 87545*

*Corresponding Author: Dr. Lorenzo Malavasi, Dipartimento di Chimica Fisica "M. Rolla", INSTM, Università di Pavia, V.le Taramelli 16, I-27100, Pavia, Italy. Tel: +39-(0)382-987921 - Fax: +39-(0)382-987575 - E-mail: lorenzo.malavasi@unipv.it

# Abstract

La$_2$Mo$_2$O$_9$ (LAMOX) is a fast oxygen ion conductor which shows high oxygen ion conductivities comparable to those of yttria-sabilized zirconia (YSZ) . LAMOX is subject to a structural phase transition from the non-conductive monoclinic form to the highly-conductive cubic form at about 580°C. The origin of the conductivity in cubic LAMOX has been suggested to be due to a "disorder" in the O sub-lattice without any insight into the real distribution of the oxygen ion. In this paper, thanks to the application of the neutron atomic pair distribution function (PDF) analysis, we provided evidences that the local structure of the cubic polymorph of LAMOX is exactly the same of that of the monoclinic phase thus indicating that the structural phase transition is actually a transition from a static to a dynamic distribution of the oxygen defects. This work represent the first application of the atomic-pair distribution function analysis to the study of an oxygen fast-oxide ion conductor and clearly indicates that a more reliable and detailed description of their local structure, particularly in the highly conductive phases, and can lead to a better comprehension of the structure-property correlation, which is the starting point for the design of new and optimized functional materials.

Fast oxygen ion conductors represent important technological materials due to their use in the field of solid state ionics and in particular as electrolytes in solid oxide fuel cells (SOFC)[1]. Currently the most commonly used materials are those based on $ZrO_2$ oxide, usually stabilized in its highly conducting cubic form by aliovalent doping such as with $Y_2O_3$[2,3]. However, the use of yttria-stabilized zirconia (YSZ) is limited by its low oxygen mobility below 1000°C. As a consequence, the need for new materials with analogous conductivities, *i.e.* in the range of $10^{-1}$ $Scm^{-1}$, at lower temperatures (600-700°C) is highly desirable in order to make them technologically useful. A strong contribution in this respect has been recently given by the discovery of a new oxygen ion conductor material of formula $La_2Mo_2O_9$, also known as LAMOX[4].

This compound exhibits a first order phase transition from the non-conductive monoclinic phase (*α*) to the highly-conductive cubic phase (*β*) at around 580°C[4] with the latter phase having a ionic conductivity of $6\times10^{-2}$ $Scm^{-1}$ at 800°C. The structure of the low temperature phase was not solved at the time of the LAMOX discovery due to its great complexity and was only solved by I.R. Evans *et al* in 2005[5]. This resulted in one of the most complex oxide structures reported up to now with 312 crystallographically independent atoms. Based on the atoms arrangement in the *α*-form it was proposed that the structure of the *β*-phase corresponds to a time average version of the room-temperature monoclinic structure[5]. The structure of the cubic form was proposed in 2000 by Lacorre[6] and then revised in 2001 by the same Authors[7]. This was determined by means of neutron diffraction and led the Authors to propose a model where two of the three available oxygen sites (O2 and O3) are partially occupied and characterized by huge atomic displacement parameters (up to $B_{eq}\sim20$ Å$^2$ for the O3 site). The presence of a strong modulation in the neutron diffraction patterns of the conducting *β*-phase of LAMOX led to the conclusion that a strong static disorder of oxygen atoms occurred in this phase[6] which was considered to be the source of the oxygen high mobility. However the use of a probe of the average structure like neutron diffraction in presence of a significant diffuse scattering will not reveal all the details of this complex compound leading to a loss of structural details which may be of great importance to understand the underlying conduction mechanism.

In order to get further insight into the structural features of the $La_2Mo_2O_9$ $\alpha$- and $\beta$-phases, test the available models for both, and try to better understand the correlation between the monoclinic and cubic polymorphs, we looked at both the short and long range order in this compound. This was accomplished by means of the atomic pair distribution function analysis (PDF) of time-of-flight neutron data collected on the $\alpha$-phase (at 500°C) and $\beta$-phase (at 600°C) of LAMOX. The power of the PDF analysis in obtaining information on the short range order in complex materials is well proved[8,9]. The strength of the technique comes from the fact that it takes all the components of the diffraction data (Bragg peaks and diffuse scattering) into account and thus reveals both the longer range atomic order and the local deviations from it. In particular, due to the significant diffuse scattering found in the LAMOX neutron patterns, we expect that a PDF analysis could help in getting a more complete picture of its structural characteristics.

Figure 1 compares a selected region of the LAMOX time-of-flight neutron patterns at 500°C (lower part) and 600°C (upper part) as a function of $Q$, the magnitude of the momentum transfer. The 500°C pattern is fully compatible with the monoclinic reference structure[5] while the 600°C pattern is in total agreement with the cubic reference structure[7]. In the inset of Figure 1 we show a very small part of the neutron patterns in order to highlight the differences in the crystal structure between the two samples. As can be seen, the monoclinic sample (lower part) presents several splittings of the analogous single-peaks in the pattern of the cubic sample (upper part) and, in addition, it is characterized by the presence of several very low-intensity superstructure peaks which are absent in the data at 600°C. Finally, in the pattern of the cubic sample (main Figure) we underlined with a red curve the peculiar undulation of the diffuse background which is peaked at around 3 Å$^{-1}$. This has been previously connected to the presence of short range order in the oxygen sub-lattice for distances around 2.5 Å[6,7], namely around the shortest O-O distance. We remark that in all the previous structural studies carried out on the LAMOX compound the information contained in the diffuse scattering was totally neglected being this peculiar background simply fitted and subtracted from the ND patterns.

From the experimental data presented above we extracted the PDFs of the monoclinic (500°C) and cubic (600°C) LAMOX through the procedure described in the Methods Section. Figure 2 (panel A) shows the two experimental PDFs overlapped in the *r*-range extending to 20 Å together with the difference curve calculated by subtracting the PDF of the 500°C data from the PDF of the 600°C data. The comparison of the data sets clearly reveals that the PDFs of the monoclinic and cubic samples are substantially identical over the range investigated. This result implies that the local structure of the two polymorphs of LAMOX is the same, as opposed to the average structures, which are markedly different (see Figure 1). Panel B indicates the expected change to the PDF if the local structure really changed from the monoclinic to the cubic phases. The two curves in Fig. 2B were calculated from the monoclinic[5,10] and cubic[6] models with all parameters the same except for the cell parameters and atomic positions. This is provided to help develop an intuitive feeling about how large the changes in the PDF would be at this phase transition if the two phases were fully ordered. Clearly, the observed changes are much smaller than the predicted.

Moreover, the use of physically reasonable thermal factors, taken from the monoclinic model, to calculate the PDF of the cubic structure leads to very sharp peaks in the PDF in contrast to the peak shape of the experimental PDF. This result is the first indication that the static disorder of the monoclinic structure, which is represented by the 216 oxygen atoms possessing different and unique atomic positions, is in a sense replaced in the cubic phase by a distribution of the oxygen ions which is locally the same as in the monoclinic phase but no longer long-range ordered. The loss of long-range order would allow the structure to become dynamic, consistent with increased ionic conductivity though our current diffraction data do not separate the static and dynamic components explicitly.

Figure 3 shows the result of the fitting procedure carried out on the experimental PDFs. Panel A shows the fit of the monoclinic model[5] to the 500°C data. The fitting parameters were the scale factor, lattice constants ($a$, $b$, $c$ and $\gamma$) and the dynamic correlation factor ($\delta$) which takes into account the correlated motion between atom pairs. Red, blue and black lines represent, respectively, the calculated PDF, the experimental PDF and the difference curve. The "goodness of fit" indicator ($R_{wp}$) of this fit is 14.7%, corresponding to a very good agreement between the experimental and calculated PDFs[11], even without refining atomic positions. In panel B we show the fit of the same monoclinic model to the 600°C data (cubic LAMOX). As can be appreciated the modelling of the experimental PDF of the cubic LAMOX by means of the monoclinic model leads to a very good agreement between the two with a $R_{wp}$ of 15.5%.

Finally, on panel C we show the result of a fit of cubic LAMOX carried out using the cubic model[7]. One of the peculiarities of this model is the presence of large atomic displacement parameters (U), particularly for the O2 ($U_{equiv}$=0.085 Å$^2$) and O3 ($U_{equiv}$=0.233 Å$^2$) sites. In order to "test" the cubic model against the experimental cubic sample in the same way as before for the monoclinic model, we refined the same parameters used in the previous fits, i.e., the scale factor, the lattice constant and the dynamic correlation factor. From the results plotted in Figure 3C, it is clear that the cubic model described in the current literature is inappropriate for describing the experimental PDF of the cubic LAMOX, particularly for $r$<10 Å. The $R_{wp}$ has now increased to 38.4%.

The inclusion of the atomic positions and atomic displacement parameters as variables in the fit of the cubic model to the 600°C data leads to a much better agreement between the experimental and calculated cubic LAMOX ($R_{wp}$ 20.0%), as it is shown in Figure 4. The final structural parameters obtained from this fit are reported in Table 1 and compared to those determined from a Rietveld refinement of the same sample, which agree with the available structural description of the cubic polymorph of LAMOX. As can be appreciated there are several significant differences between the two data sets, particularly for the Mo coordinate, the *y*-coordinate of O2 and the overall position of the O3. Finally, with the only exception of the Mo, the atomic displacement parameters of all the atoms are all lower, in particular for the O3 atom, even though all of them remain quite high. We note that a possible extension of this model to also include the refinement of the atoms occupancies leads to meaningless results for the O1 position (always bigger that 1.2). More reliable results with reference to the occupancies may be obtained by extending the fit of the PDF to a higher *r*-range. In this case the final fit results are closer to those obtained from the Rietveld refinement, particularly for the atomic displacement parameters and the occupancies, since we are now looking at the average structure of the sample.

The results presented so far have shown that the average structures for the LAMOX compound at 500°C and at 600°C are different and, in particular, monoclinic at the first temperature and cubic at the second. However, a striking feature of the PDFs of the two phases is that by crossing the phase transition the local structure does not substantially changes, as exemplarily shown in Figure 2, panel A. These results strongly support a scenario in which the structural phase transition from the *α* to the *β*-phase is a transition from a long-range ordered to a, presumably dynamic, short-range ordered distribution of the oxygen defects while preserving the monoclinic local structure. In particular, the monoclinic phase presents a Mo-O environment of variable coordination, namely 4, 5 and 6, which, by considering the oxygen sharing pattern become 3.73, 4.77 and 4.94[5]. In the cubic phase the "average" Mo-O coordination number is 4.5. The peculiar nature of the $\alpha \rightarrow \beta$ phase transition is the local change (with time) of the Mo-O coordination environment due to the migration of the oxygen ions through a vacancy mechanism. An average probe such as the neutron diffraction well describes the long-range static distribution of the oxygen ions in the monoclinic phase, but fails in describing the dynamic distribution in the cubic phase. In this last case all this information is found in the diffuse scattering due to the short-range order which is clearly visible in the neutron pattern of the sample at 600°C. Since this information is not used in the traditional Rietveld refinement of the neutron diffraction patterns the only way to deal with the oxygen sub-lattice is by introducing large values of the atomic displacement parameters for the oxygen ions, as done in the available models describing the cubic structure by means of ND[6,7]. The consequence on the calculated PDF of this huge thermal factors is a smearing of the oxygen atom density leading to a poor description of the local structure (see Figure 3, panel C).

The power of the PDF analysis, which makes use of the total scattering function, is nicely probed in the present case. The PDFs of LAMOX at 500°C and 600°C are substantially identical showing that on the *local scale* the structure, in both cases, is monoclinic. The use of the cubic model to refine data at 600°C has been proved to be possible (see Fig. 4) but, even in the case of a real-space refinement including all the scattering components, the final result is a cubic structure where the large atomic displacement parameters compensate for the dynamic distribution of the atoms and the inability to precisely define their distribution. However, it is clear by comparing Figure 3 (panel C) and Figure 4 that the total scattering method allows a better description of the cubic sample with the cubic model, even though is also ends, like in the Rietveld refinement, with high thermal factors which are not simply due to the dynamics of the oxygen migration but to some deficiencies in the model. The cubic model does not correctly capture the geometry of the oxygen ion coordination and is lacking in describing the O-occupancies on the short-range. This can be understood if we compare the $U$ values of LAMOX with others fast oxygen-ion conductors such as YSZ. For example, a time-of-flight study of YSZ[12] at 767°C determined a $U_{iso}$ for the O atoms of 0.018 Å$^2$. Even for higher temperatures the $U$ values are always significantly lower than those of O atoms in LAMOX. It should be noted that the highest a.d.p. pertains to the Mo which is the ion most involved in the oxygen diffusion through a change in its coordination environment.

The main result presented so far, i.e. the equivalence of the local structure in the monoclinic and cubic phase, is of great importance since it demonstrates that what is usually considered as a "disordered oxygen distribution" has actually a well defined local structure which is the same as in the monoclinic phase but dynamic. As has been suggested by Evans and co-workers, the structure of the oxide ion conducting phase corresponds to a time-averaged version of the monoclinic phase[5]. However, in the present work we took a step further in elucidating the correlation between the $\alpha$ and $\beta$-phase of LAMOX by *directly* showing this relationship through the PDF analysis. This understanding allows us to consider different possible oxygen ion conduction paths in the conducting $\beta$-phase of LAMOX by considering, as a starting point, the monoclinic structure. A possible proposal of this conduction mechanism may be done on the basis of the Mo-O coordination environment as it is found in the monoclinic phase. It is plausible that the conduction mechanism involves the migration of an oxygen ion from a Mo-O polyhedron with a higher coordination number towards an oxygen vacancy on another Mo-O polyhedron. Interestingly, it is possible to follow a path in the monoclinic structure which involves distances from an O belonging to a first "donor" Mo-O polyhedron to the Mo of another "acceptor" Mo-O polyhedron of about 5.6 Å (Mo-O- V$_O$-Mo). This distance is exactly the one separating two Mo belonging to two different polyhedra that are connected by two oxygen atoms each belonging to a separate polyhedron and with a partial O-occupancy, i.e., for example, Mo-O2-O3-Mo. This result suggests that when the LAMOX transforms from the $\alpha$ to the $\beta$-phase the oxygen start "jumping" from a Mo-O group to the next Mo-O group leading to, on average, two sites with partial occupancy (what is seen from the average probe) that are actually sites within a polyhedron with a well defined coordination which is changing with time.

In conclusion, we have provided the first application of the atomic-pair distribution function analysis to the study of an oxygen fast-oxide ion conductor of actual interest in the SOFC community. Our results have shown that a clear and reliable description of the local atom arrangement in LAMOX structure can be only achieved through the application of a local probe such as the PDF. This allowed us to *directly* determine that the transition from the monoclinic to the cubic phase of LAMOX is a transition from a static to a dynamic distribution of the oxygen defects while preserving the monoclinic local structure. The application of the PDF analysis to the solid state ionics materials can allow a more detailed description of their local structure, particularly in the highly conductive phases, and leads to a better comprehension of the structure-property correlation, which is the starting point for the design of new and optimized functional materials.

# Experimental

$La_2Mo_2O_9$ has been prepared by conventional solid state reaction by mixing stoichiometric amounts of high purity $La_2O_3$ and $MoO_3$ powders (Aldrich > 99.9 %) into ethanol in an agate mortar, compressed and then heated at 500°C for 24 hours, then at 700°C for others 24 hours and finally at 900°C for 60 hours, during which the sample has been re-ground twice. Phase purity was checked by means of room temperature x-ray powder diffraction (XRPD) patterns acquired with a Bruker D8 Advance diffractometer.

Neutron powder diffraction measurements were carried out on the NPDF[14] diffractometer at the Lujan Center at Los Alamos National Laboratory. 10g of LAMOX powder sample was packed in a cylindrical vanadium tube and measured from 300 to 750K. Data were processed to obtain the PDFs using PDFgetN[13] and the structural modelling was carried out using the program PDFFIT2[11].

## Acknowledgements

Work in the Billinge group was supported by grant NSF DMR-0304391. We would like to acknowledge helpful discussions with Emil Bozin and Ahmad Masadeh. This work has benefited from the use of NPDF at the Lujan Center at Los Alamos Neutron Science Center, funded by DOE Office of Basic Energy Sciences. Los Alamos National Laboratory is operated by Los Alamos National Security LLC under DOE Contract DE-AC52-06NA25396. The upgrade of NPDF has been funded by NSF through grant DMR 00-76488.

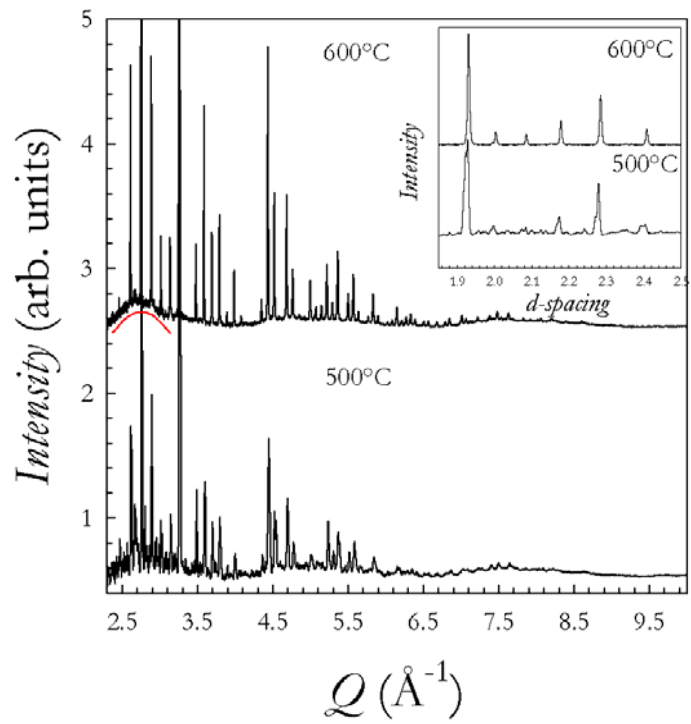

**Figure 1** –Time-of-flight neutron diffraction patterns of La$_2$Mo$_2$O$_9$ at 500°C (lower part) and 600°C (upper part). The inset shows a small part of the pattern on an expanded scale for both temepratures for comparison.

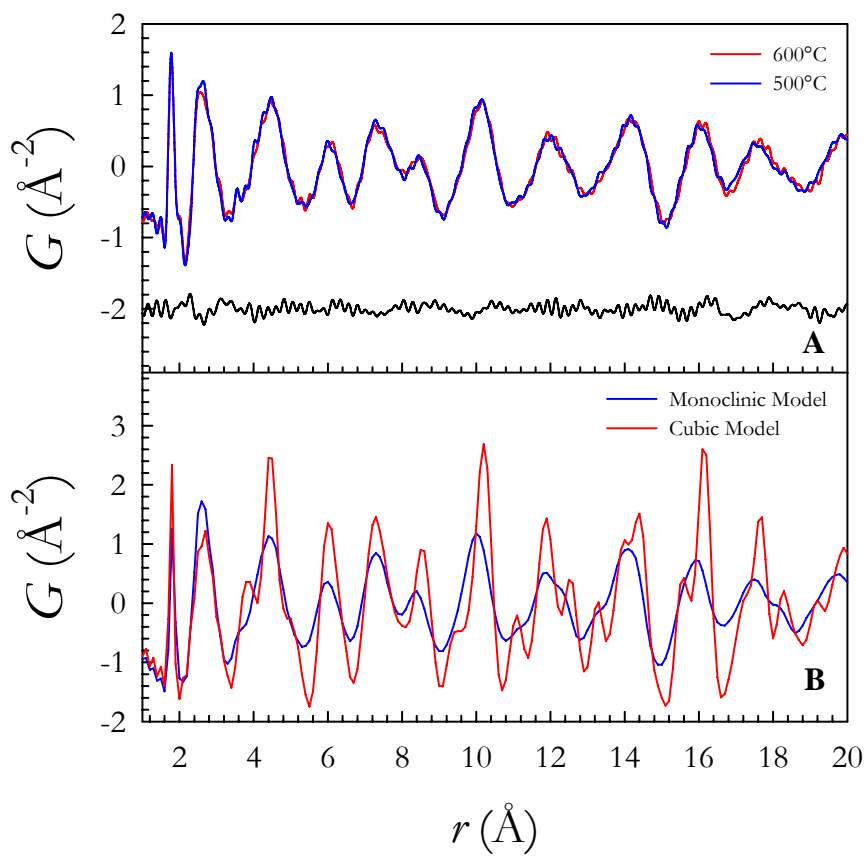

**Figure 2 –** **A:** comparison between the experimental PDFs of LAMOX at 500°C (blue line) and 600°C (red line) and their difference (black line). **Panel B**: comparison between the calculated PDFs of LAMOX from the monoclinic (blue line) and cubic (red line) models. Details about the calculated PDFs are in the text.

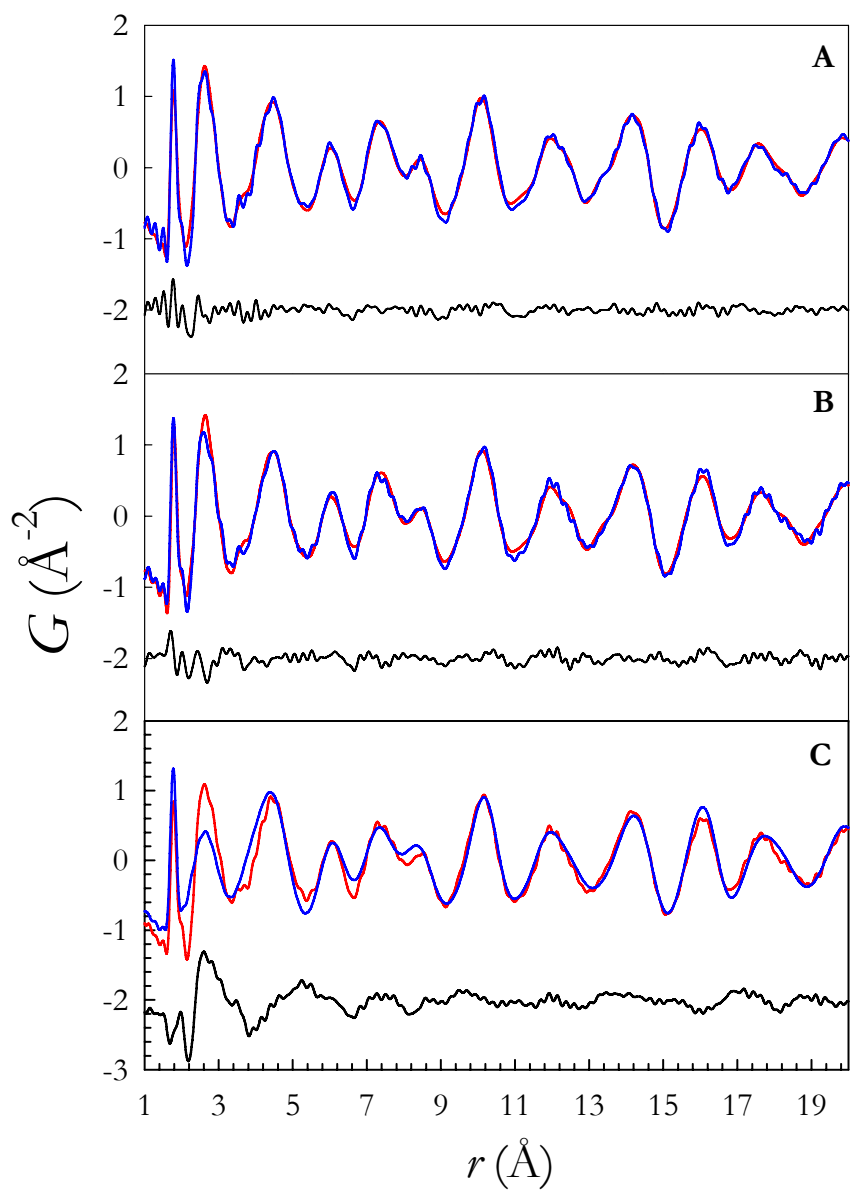

**Figure 3 –Panel A:** fit of the experimental PDF of LAMOX at 500°C with the monoclinic model (red, blue and black lines represent, respectively, the calculated PDF, the experimental PDF and the difference curve in all the three panels). **Panel B**: fit of the experimental PDF of LAMOX at 600°C with the monoclinic model. **Panel C**: fit of the experimental PDF of LAMOX at 600°C with the cubic model. In all the refinements the starting model was from the published structure in the literature[5,6] and the parameters fitted in the refinement where: scale factor, dynamic correlation factor and lattice constants.

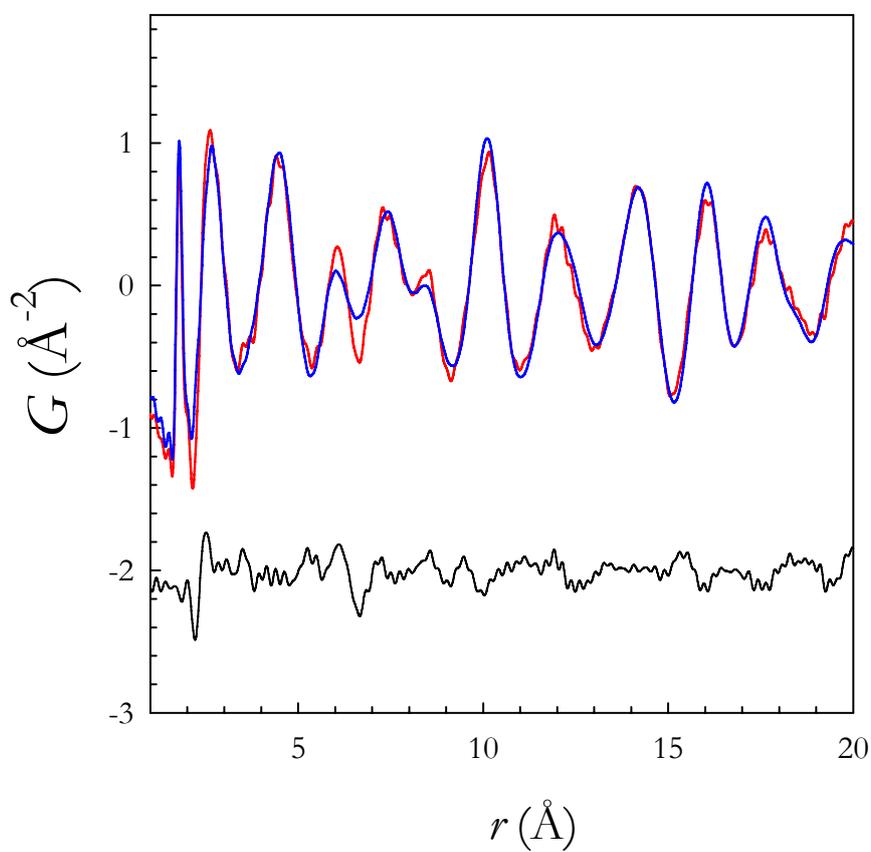

**Figure 4** –Fit of the experimental PDF of LAMOX at 600°C with the cubic model (red, blue and black lines represent, respectively, the calculated PDF, the experimental PDF and the difference curve in all the three panels). The fit was carried out in the same way as those shown in Fig. 3, but the atomic positions and atomic displacement parameters were allowed to vary.

**Table 1 -** Fit results for the Rietveld refinement of the cubic LAMOX and for the PDF fit represented in Figure 4, i.e. carried out by using the cubic model.

| Parameter | Rietveld Refinement | Fit of Figure 4 |
|---|---|---|
| *Lattice parameter* | 7.21788(8) | 7.2220(3) |
| La, *x* | 0.8543(3) | 0.8575(2) |
| $U_{iso}$ | 0.065(7) | 0.1004(8) |
| Mo, *x* | 0.1682(2) | 0.1754(2) |
| $U_{iso}$ | 0.057(6) | 0.0750(4) |
| O1, *x* | 0.3147(4) | 0.3103(2) |
| $U_{iso}$ | 0.093(3) | 0.0578(6) |
| O2, *x* | 0.9933(5) | 0.9859(2) |
| O2, *y* | 0.1875(9) | 0.1615(3) |
| O2, *z* | 0.3438(5) | 0.3455(2) |
| $U_{iso}$ | 0.088(4) | 0.0614(5) |
| O3, *x* | 0.894(1) | 0.9830(3) |
| O3, *y* | 0.698(6) | 0.7531(3) |
| O3, *z* | 0.553(1) | 0.4950(3) |
| $U_{iso}$ | 0.33(3) | 0.0310(7) |
| $R_{wp}$ (%) | 2.99 | 18.7 |

*TABLE OF CONTENTS (TOC)*

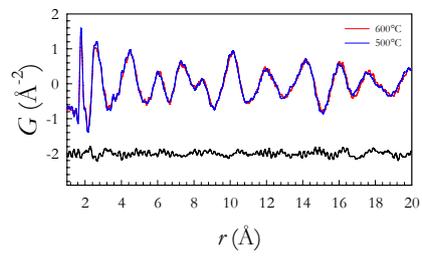

Comparison between the experimental PDFs of LAMOX at 500°C (blue line) and 600°C (red line) and their difference (black line).